\documentclass[12pt]{article}
\usepackage{graphicx,color}
\begin{document}
\baselineskip=5.5mm
\newcommand{\be} {\begin{equation}}
\newcommand{\ee} {\end{equation}}
\newcommand{\Be} {\begin{eqnarray}}
\newcommand{\Ee} {\end{eqnarray}}
\renewcommand{\thefootnote}{\fnsymbol{footnote}}
\def\a{\alpha}
\def\b{\beta}
\def\g{\gamma}
\def\G{\Gamma}
\def\d{\delta}
\def\D{\Delta}
\def\e{\epsilon}
\def\k{\kappa}
\def\l{\lambda}
\def\L{\Lambda}
\def\t{\tau}
\def\om{\omega}
\def\Om{\Omega}
\def\s{\sigma}
\def\lg{\langle}
\def\rg{\rangle}
\def\G{{\bf G}}
\newcommand{\tblue}[1]{\textcolor{blue}{#1}}
\begin{center}
{\large {\bf Nonlinear response theory for Markov processes: 
Simple models for glassy relaxation} }\\
\vspace{0.5cm}
\noindent
{\bf Gregor Diezemann} \\
{\it
Institut f\"ur Physikalische Chemie, Universit\"at Mainz,
Jakob-Welder-Weg 11,\\ 55128 Mainz, FRG\\

\vspace{0.25cm}
08.03.2012\\
\vspace{0.25cm}
}
\end{center}
\vspace{0.75cm}
\vspace{0.75cm}

\noindent
{\it
The theory of nonlinear response for Markov processes obeying a master equation is formulated in terms of time-dependent perturbation theory for the Green's functions and general expressions for the response functions up to third order in the external field are given.
The nonlinear response is calculated for a model of dipole reorientations in an asymmetric double well potential, a standard model in the field of dielectric spectroscopy.
The static nonlinear response is finite with the exception of a certain temperature $T_0$ determined by the value of the asymmetry.
In a narrow temperature range around $T_0$, the modulus of the frequency-dependent cubic response shows a peak at a frequency on the order of the relaxation rate and it vanishes for both, low frequencies and high frequencies.
At temperatures at which the static response is finite (lower and higher than $T_0$), the modulus is found to decay monotonously from the static limit to zero at high frequencies.
In addition, results of calculations for a trap model with a Gaussian density of states are presented. 
In this case, the cubic response depends on the specific dynamical variable considered and also on the way the external field is coupled to the kinetics of the model. 
In particular, a set of different dynamical variables is considered that gives rise to identical shapes of the linear susceptibility and only to different temperature dependencies of the relaxation times.
It is found that the frequency dependence of the nonlinear response functions, however, strongly depends on the particular choice of the variables.
The results are discussed in the context of recent theoretical and experimental findings regarding the nonlinear response of supercooled liquids and glasses.
}

\vspace{1cm}
\section*{I. Introduction}
In recent years progress has been achieved in the understanding of the heterogeneous dynamics observed in supercooled liquids and glassy systems\cite{Sillescu99, Berthier:2011p6852}.
Starting with NMR experiments\cite{SRS91, HWZS95, G13} a number of frequency-selective techniques have been developed in order to investigate the nature of the dynamic heterogeneities in the slow primary relaxation of supercooled liquids\cite{G23, Ediger00, Israeloff00, Richert02}.
Also the length scale associated with the heterogeneities could be determined in some 
cases\cite{Tracht98, Reinsberg01}.
In the experimental studies the system always is monitored at more than two times via the observation of four-time correlation functions as in the quoted NMR experiments. 
Alternatively, large external fields are applied giving rise to nonlinear effects as in the nonresonant hole burning studies\cite{SBLC96, G16}.
Furthermore, in computer simulations on model systems dynamic heterogeneities have been 
observed via following certain trajectories\cite{Kob97, Doliwa98} or also via the calculation of four-time correlation functions\cite{Schroder03, Reichman07}. 

Most of the studies on dynamic heterogeneities were concerned with systems in thermal equilibrium, but also aging glasses have been investigated\cite{Ediger02, Lunki05}. 
Heterogeneous aging has also been studied theoretically in spin glasses\cite{Castillo02}, in simple spin models\cite{G48} and also in a free-energy landscape model for glassy relaxation\cite{G56}.

In recent years, both experimental techniques and theoretical tools have been refined in order to allow detailed investigations of dynamic heterogeneities.
In particular, it has been recognized that higher-order correlation functions that probe the system at different times and different locations in space can be used to observe a length scale\cite{Berthier05} and the relevant four-point correlation function $\chi_4(t)$ has been studied 
theoretically\cite{Toninelli05, Berthier07a, Berthier07b}.
Earlier experimental studies used the approximative relation of $\chi_4(t)$ to a two-point correlation 
function\cite{Berthier05, DalleFerrier07} in order to extract the number of cooperatively rearranging particles, $N_{\rm corr}$.
In an influential paper Bouchaud and Biroli related the nonlinear (cubic) response $\chi_3(\om,T)$ to 
$\chi_4(t)$\cite{Bouchaud05}.
The experimental determination of $\chi_3(\om,T)$ allowed the determination $N_{\rm corr}$ more 
directly\cite{CrausteThibierge10, Brun11} and the results are compatible with the earlier observations.
In particular, it was argued that the function
\be\label{X3.Def}
X(\om,T)=\left|\chi_3(\om,T)\right|{k_BT\over(\D\chi_1)^2a^3}
\ee
with $\D\chi_1$ denoting the static linear response, $k_B$ the Boltzmann constant and $a^3$ the molecular volume, exhibits a hump-like structure.
This behavior is assumed to be a distinctive feature of glassy correlations\cite{CrausteThibierge10}.
Additionally, the maximum of $X(\om,T)$ is expected to decrease with increasing temperature and to be directly proportional to $N_{\rm corr}$.
If glassy correlations are absent, $X(\om,T)$ should not be peaked and this 'trivial' behavior consists in a smooth cross-over from a low-frequency limiting value to a vanishing high-frequency limit.
In this context it has to be mentioned that Brun et al. found a hump-like shape for $X(\om,T)$ in a calculation employing the so-called box model\cite{Brun11b}, a model devoid of spatial aspects.

Apart from the determination of $N_{\rm corr.}$ the nonlinear dielectric response has been used to investigate the nature of the heterogenous dynamics via comparison of the cubic response with the linear 
response\cite{Richert06, Weinstein07} and the results were discussed in the framework of the box model.
Similar measurements were performed in order to extract the configurational heat capacity of 
liquids\cite{Wang07}.
In addition also the nonlinear dielectric response of liquids due to an AC and an DC field pulse have been 
recorded\cite{Rzoska12} and also dipolar glasses have been investigated\cite{Roland98}.

The present paper deals with the theory of nonlinear response functions for Markov processes, because the  relaxation in complex systems often is modeled in terms of such stochastic dynamics.
For systems that follow a Hamiltonian or Langevin dynamics, nonlinear response functions have been considered quite some time ago\cite{BC59, APR76, Morita86}.
However, explicit calculations of response functions are rare and most of them relate to variants of the rotational diffusion of molecules in the presence of strong electric fields, see e.g. 
refs.\cite{DD95, Dejardin00, Kalmykov01, G57}.
In addition, approximate nonlinear response theory has been investigated more generally\cite{Dyre89} and also fluctuation-dissipation relations beyond the linear regime have been discussed\cite{Eyink00, Lippiello08}.
The nonlinear response of supercooled liquids has been worked out theoretically in the framework of mode-coupling theory\cite{Tarzia10}.
Here, I will perform the calculation of the response functions in close analogy to the quantum-mechanical way of computing response functions\cite{muk95}.
Time-dependent perturbation theory for the propagator is used in order to obtain the response in the desired order in the amplitude of the external field.
I will present the results of calculations of the cubic response function for two Markovian models of relaxation.
One model describes the reorientations of dipoles in an asymmetric double well potential (ADWP) and has been used to interpret results of dielectric experiments in general\cite{Frohlich49}.
Furthermore, it has also been employed in calculations of the signals obtained in nonresonant holeburing experiments\cite{G46}.
It will be shown that $X(\om,T)$ mainly behaves 'trivially' for this model.
Another model that will be considered is the trap model with a Gaussian density of states\cite{Dyre95, MB96}.
This model has been used in the interpretation of some features of the relaxation in simulated supercooled 
liquids, both in equilibrium\cite{Denny03} and in the aging regime\cite{G64, G71}. 
Here, the results for $X(\om,T)$ are more complex and, depending on the parameters chosen, either exhibit a peak-like structure or 'trivial' behavior.

The paper is organized as follows. 
In the next section, I will outline the calculation of nonlinear response functions for systems obeying a master equation. 
For convenience of the reader, most of the explicit calculations are presented in the Appendix.
The sections following this theoretical part deal with a discussion of the results obtained for the two models considered and the paper closes with some concluding remarks.
\section*{II. Nonlinear response theory for Markov processes}
In this section, I will outline the general procedure to calculate the nonlinear response functions for a system that is described by a master equation (ME)\cite{vkamp81, Gardiner97}.
If one is dealing with complex systems a coarse-grained procedure may result in a description of the underlying dynamics in terms of a non-stationary Markov process. 
Therefore, in order to keep the treatment general, I will treat the case of a ME with time-dependent transition rates.

In the following, $G_{kl}(t,t_0)$ denotes the conditional probability to find the system in state $k$ at time $t$ provided it was in state $l$ at time $t_0$ (Green's function, propagator) in a discrete notation.
If continuous variables are considered, all sums in the following expressions are to be replaced by the corresponding integrals. 
Denoting the rates for a transition from state $k$ to state $l$ by $W_{lk}(t)$, the ME reads:
\be\label{ME.t.abh}
{\partial\over\partial t}G_{kl}(t,t_0)=
-\sum_nW_{nk}(t)G_{kl}(t,t_0)+\sum_nW_{kn}(t)G_{nl}(t,t_0)
\ee
This equation has to be solved with the initial condition $G_{kl}(t_0,t_0)\!=\!\d_{kl}$, where $\d_{kl}$ denotes the Kronecker symbol.
If the transition rates $W_{kl}(t)$ are time-independent the process considered is stationary.
The one-time probabilities $p_k(t)$ (the populations of the states) obey the same ME and are given by
$p_k(t)=\sum_lG_{kl}(t,t_0)p_l(t_0)$.
The $W_{kl}(t)$ can be related to the elements of the master-operator ${\cal W}(t)$ via\cite{vkamp81}:
\be\label{We.kl.t}
{\cal W}(t)_{kl}=W_{kl}(t)-\d_{kl}\sum_nW_{nl}(t)
\ee
Here, ${\cal W}(t)_{kl}\geq 0$ holds for all $k\neq l$ and the sum rule $\sum_k{\cal W}(t)_{kl}=0$
is fulfilled for all values of $l$ as it is a general property of the transition rates for any Markov process.
At the initial time $t_0$ the system is described by a fixed set of populations,
$p_k^0\!=\!p_k(t_0)$ with $\sum_kp_k^0\!=\!1$.
If a stationary system is considered, one often starts from equilibrium populations $p_k^0=p_k^{\rm eq}$ or if one is interested in describing a situation with a certain thermal history one might choose the $p_k^0$ as the equilibrium populations at a temperature different from the working temperature.

In order to treat the system in the presence of an external field one has to specify the field-dependence of the transition rates, which is not straightforward.
In case of Hamiltonian or Langevin dynamics, the linear coupling of a variable $M(t)$ to a field $H(t)$ gives rise to an extra term $[-M(t)\cdot H(t)]$ in the Hamiltonian.
In a Fokker-Planck equation, this gives rise to a term linear in $H$\cite{Risken89}.
If one considers a ME, one choice that has been used in a number of investigations of fluctuation-disspiation relations is given by
\be\label{Wkl.HX}
W_{kl}^{(H)}(t)=W_{kl}(t)e^{\b H[\g M_k-\mu M_l]}
\ee
with arbitrary $\g$ and $\mu$\cite{CR03, G39, G54}.
In this expression $\b=T^{-1}$ denotes the inverse temperature with the Boltzmann constant set to unity, $k_B=1$.
If the system obeys detailed balance, one has the restriction $\g+\mu=1$.
In particular, for systems described by a Fokker-Planck equation, one would naturally choose $\g=\mu=1/2$ and a linear expansion of eq.(\ref{Wkl.HX}) gives the usual term in the Fokker-Planck operator.
However, it is obvious from eq.(\ref{Wkl.HX}) that in general one will have nonlinear contributions to the perturbation also if the coupling to the field is linear in the sense described above.
This means that couplings of a form like $[\tilde M(t)\cdot H^2]$, as it would appear for instance if the coupling to an induced dipole-moment is considered\cite{DD95}, are absent.

In order to keep the treatment general, I will formulate the response theory without fixing the field-dependence of the transition rates.
It is only assumed that it can be cast in the form:
\be\label{W.H.kl.Taylor}
W_{kl}^{(H)}(t)=\sum_{n=0}^\infty{1\over n!}W_{kl}^{(n)}(t)\cdot[\b H(t)]^n
\quad\mbox{with}\left.\quad W_{kl}^{(n)}(t)={d^n\over d(\b H)^n}W_{kl}^{(H)}(t)\right|_{H=0}
\ee
The elements of the propagator $\G^{(H)}(t,t_0)$ are obtained from the ME, eq.(\ref{ME.t.abh}), where the field-independent quantities are replaced by those explicitly depending on the external field, i.e.
$\dot G^{(H)}_{kl}(t,t_0)=-\sum_nW^{(H)}_{nk}(t)G^{(H)}_{kl}(t,t_0)+\sum_nW^{(H)}_{kn}(t)G^{(H)}_{nl}(t,t_0)$.
The solution of this equation is needed to calculate the response of the system to an external field applied at time $t_0$ and measured by an observable $F(t)$, 
\be\label{F.expect}
\lg F(t)\rg_{(H)}=\sum_{kl}F_kG^{(H)}_{kl}(t,t_0)p_k(t_0)
\ee
In order to be able to set up a perturbation theory for $\G^{(H)}(t,t_0)$ in terms of the corresponding 'field-free' propagator $\G(t,t_0)$, one uses the decomposition
\be\label{W.H.Vn}
{\cal W}^{(H)}(t)={\cal W}(t)+{\cal V}(t)
\quad\mbox{with}\quad
{\cal V}(t)=\sum_{n=1}^\infty{\cal V}^{(n)}(t)
\ee
where the perturbation is given according to eq.(\ref{W.H.kl.Taylor})
\be\label{Vn.def}
{\cal V}^{(n)}(t)_{kl}={[\b H(t)]^n\over n!}\left[W_{kl}^{(n)}(t)-\d_{kl}\sum_nW_{nl}^{(n)}(t)\right]
\ee
The theoretical treatment is very similar to the one utilized in ref.\cite{G54} and consists in performing time-dependent perturbation theory to treat ${\cal V}(t)$ in the desired order of the field.
The details of this procedure are described in Appendix A. 
The explicit expressions for the response functions are given up to third order in the field and the extention to higher order is straightforward.

\begin{figure}[h!]
\centering
\includegraphics[width=7.5cm]{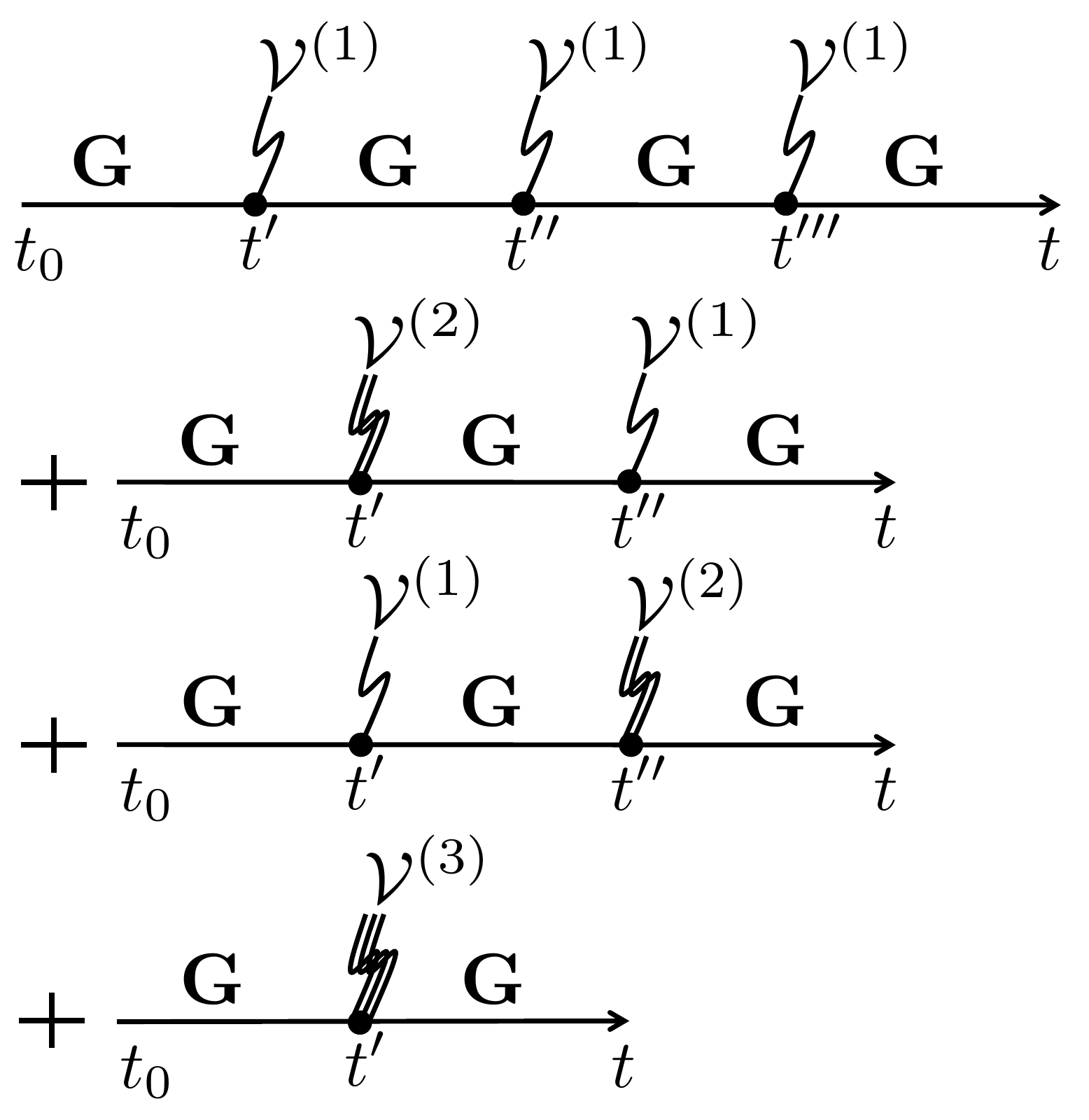}
\vspace{-0.5cm}
\caption{Pictorial representation of the perturbation expansion for the third-order response. The unperturbed propagators are denoted by $\G$ and the ${\cal V}^{(n)}$ are the perturbations according to 
eq.(\ref{Vn.def}). }
\label{Plot1}
\end{figure}

The main difference to the formalism utilized for Hamiltonian or Langevin dynamics with a linear coupling to the external field is that here in general the elements ${\cal V}^{(n)}(t)_{kl}$ with $n>1$ do not vanish.
This gives rise to a number of extra terms.
The situation is visualized in Fig.\ref{Plot1}, which shows the diagrams representing the interaction with the field for the third-order response.
One has the terms stemming from purely linear interactions given in the first line. 
These terms also appear in a Fokker-Planck treatment of a linear coupling.
Furthermore, one has two cross terms between first-order and second-order perturbations (second and third line in Fig.\ref{Plot1}) and a term stemming from the third-order perturbation (fourth line).
For Langevin dynamics, cross-terms only appear if a quadratic coupling is considered in addition to a linear one.

While in Appendix A the general expressions for the response functions are given, in the actual model calculations I will consider only the response of systems that are in thermal equilibrium prior to the application of the external field.
Furthermore, the models treated in the present paper represent stationary Markov processes with time-independent transition rates.
The discussion will be limited to sinusoidal fields of the form
\be\label{H.cos.om.t}
H(t)=H_0\cos{(\om t)}
\ee
For this oscillating field the linear and the cubic response for times long compared to the initial transients
can be written as:
\Be\label{Chi.om.def}
\chi^{(1)}(t)
&&\hspace{-0.6cm}=
{H_0\over2}\left[e^{-i\om t}\chi_1(\om)+c.c.\right]
\nonumber\\
\chi^{(3)}(t)
&&\hspace{-0.6cm}=
{H_0^3\over2}\left[e^{-i\om t}\chi_3^{(1)}(\om)+e^{-i3\om t}\chi_3^{(3)}(\om)+c.c.\right]
\Ee
where $c.c.$ denotes the complex conjugate.

In the following sections, I will mainly discuss the quantity $X(\om,T)$ introduced in eq.(\ref{X3.Def}).
As the models that will be considered in the following are not related to any spatial aspects of dipole reorientations or relaxing units, the molecular volume will be set to unity, $a^3=1$.
Additionally, one has a separate function for each frequency-component, cf. ref.\cite{Brun11},
that can be written as ($\a=1, 3$):
\be\label{Xalfa.Def}
X_\a(\om,T)={T\over(\D\chi_1)^2}\left|\chi_3^{(\a)}(\om,T)\right|
\ee
This function eliminates the 'trivial' temperature dependence of $\chi_3^{(\a)}(\om,T)$ because 
$\D\chi_1\sim\b$ , cf. eq.(\ref{Chi1}) and $\chi_3^{(\a)}\sim\b^3$ according to eq.(\ref{Chi3}).
Therefore, any temperature dependence stems from the 'intrinsic' relaxation behavior of the dynamical variable considered.
\section*{III. The ADWP-model for dipole reorientations}
In this section, I will present the results for one of the simplest models for dielectric relaxation, namely the model of dipole reorientation in an asymmetric double well potential.
I will closely follow the notation used in a related investigation of the nonresonant dielectric hole burning technique\cite{SBLC96, G16, G46}.

As in ref.\cite{G46}, two dipole orientations denoted by '$1$' and '$2$', characterized by polar angles 
$\theta_1=\theta$ and $\theta_2=\theta+\pi$ are assumed and the transition rates between the two are given by
$W_{12}=We^{-\b\D/2}$ and $W_{21}=We^{+\b\D/2}$. 
Here $\D$ denotes the asymmetry, and $W$ is the hopping rate in the symmetric case.
For this model, the Green's functions in the field-free case are are given by:
\be\label{Gkl.ADWP}
G_{kl}(t)=p_k^{\rm eq}\left(1-e^{-t/\t}\right)+\d_{kl}e^{-t/\t}
\quad\mbox{with}\quad \t^{-1}=2W\cosh{\!(\b\D/2)}
\quad\mbox{and}\quad p_k^{\rm eq}=\t\cdot W_{kl} 
\ee
The variable that couples to the field is 
\[
M_k=M\cos{\!(\theta_k)}
\quad\mbox{and therefore}\quad
M_1=M\cos{\!(\theta)}
\quad;\quad
M_2=-M\cos{\!(\theta)}
\]
with $M$ denoting the static molecular dipole moment.
The field-dependent transition rates are chosen as in eq.(\ref{Wkl.HX}) with $\g=\mu=1/2$.
(If this restriction is relaxed all response functions depend on the sum $(\g+\mu)$, which equals unity in the present case.)
In the calculation of the response I assume a collection of systems characterized by an isotropic distribution of orientations and therefore an average over the angle $\theta$ is performed according to
$\lg\cos^n{\!(\theta_k)}\rg=(n+1)^{-1}$ for $n$ even and $\lg\cos^n{\!(\theta_k)}\rg=0$ for $n$ odd.

Using the general expressions given in Appendix A along with eq.(\ref{Gkl.ADWP}), one finds for the linear response:
\be\label{Chi1.ADWP}
\chi_1(\om)=\D\chi_1{1\over1-i\om\t}
\quad\mbox{where}\quad
\D\chi_1=\b\lg \D M^2\rg
=\b{M^2\over3}\left(1-\d^2\right)
\ee
In this expression, I defined $\d=\tanh{\!(\b\D/2)}$. (It should be mentioned that $\D\chi_1$ differs by a factor $1/2$ from the definition of $\chi_{DWP}$ in ref.\cite{G46}.) 
As usual, $\D\chi_1$ is related to the mean-square fluctuations of the dipole moment
$\lg \D M^2\rg$.
Eq.(\ref{Chi1.ADWP}) follows immediately from the definition $\lg M^m\rg=\sum_kM_k^mp_k^{\rm eq}$ and 
eq.(\ref{Gkl.ADWP}) with additional isotropic average.

Note that in the ADWP-model, the static susceptibility $\D\chi_1$ for non-vanishing asymmetry depends on temperature due to the dependence on $\d$ in addition to the trivial $1/T$-dependence. 
This behavior for finite asymmetry is different from the model of Brownian rotational diffusion, where 
$T\D\chi_1$ is independent of temperature\cite{Frohlich49}. 
For vanishing asymmetry, the models show identical behavior (apart from irrelevant prefactors).
Without showing results here, it is mentioned that $\rm Re(\chi_1(\om))$ decays from its low-frequency limit $\D\chi_1$ to zero for large frequencies and $\rm Im(\chi_1(\om))$ shows the typical Lorentzian behavior and is peaked at $\om\t=1$.

\begin{figure}[h!]
\centering
\includegraphics[width=8.0cm]{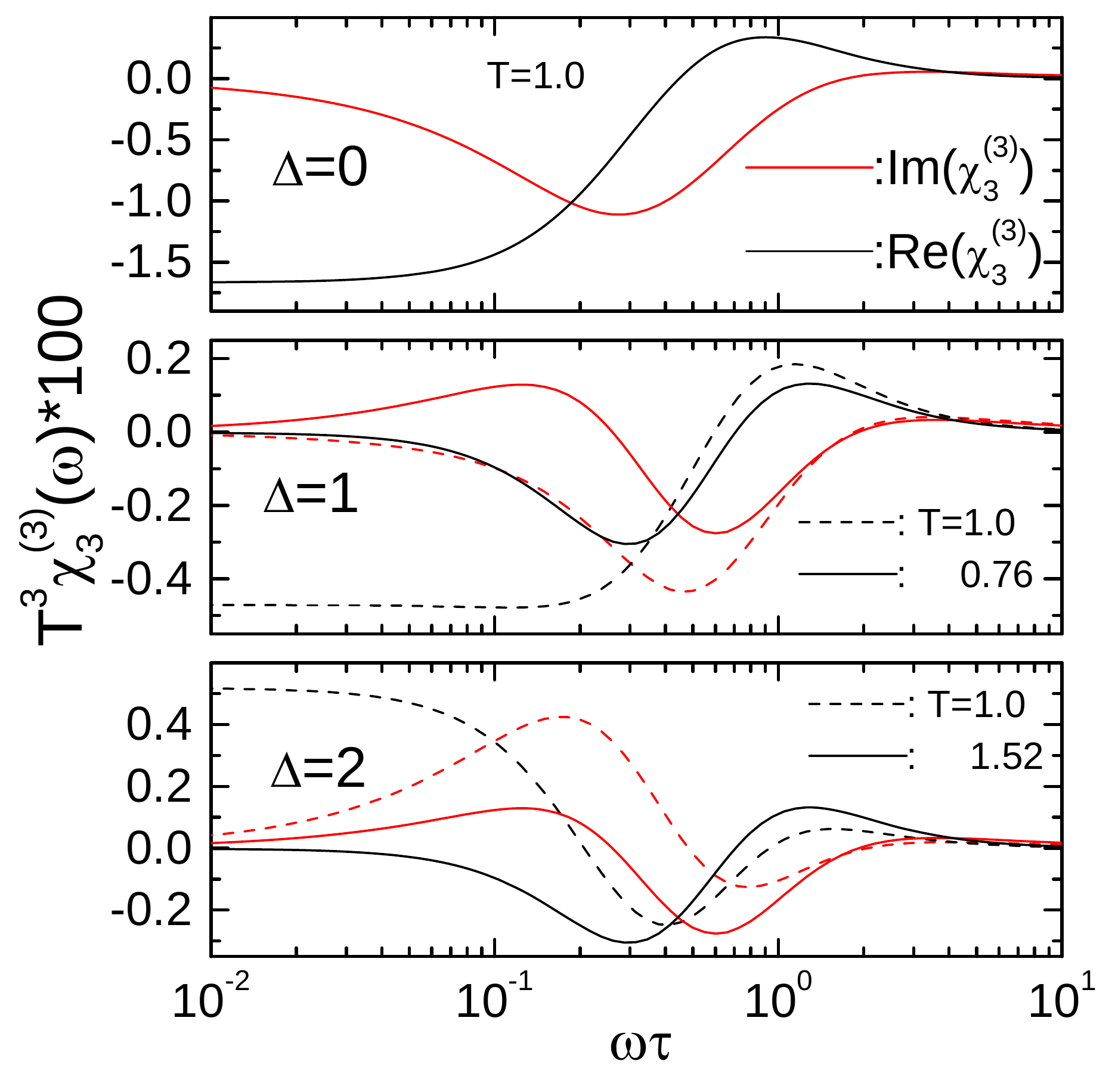}
\vspace{-0.5cm}
\caption{Real part (red) and imaginary part (black) of the 3$\om$-component $\chi_3^{(3)}(\om)$ for the ADWP-model as a function of $\om\t$, where $\t$ is the relaxation time according to eq.(\ref{Gkl.ADWP}). }
\label{Plot2}
\end{figure}

The third-order response functions are calculated according to eq.(\ref{Chi.om.def}) using the general expressions given in eq.(\ref{Chi3}) in the Appendix.
In a straightforward calculation one finds:
\be\label{Chi3.ADWP}
\chi_3^{(\a)}(\om)={M^4\over20}\b^3\left(1-\d^2\right)\times S_3^{(\a)}(\om\t)
\ee
Here, the spectral functions only depend on the product $x=\om\t$ and are given by:
\Be\label{S3n.ADWP}
S_3^{(1)}(x)
&&\hspace{-0.6cm}=
\d^2{3(1+i2x)\over(1+x^2)(1+4x^2)}+{2(x^2-1)+ix(x^2-3)\over2(1+x^2)^2}
\\
S_3^{(3)}(x)
&&\hspace{-0.6cm}=
\d^2{(1-11x^2)+i6x(1-x^2)\over(1+x^2)(1+4x^2)(1+9x^2)}
+{2(5x^2-1)+i3x(x^2-3)\over6(1+x^2)(1+9x^2)}
\nonumber
\Ee
When compared to the model of Brownian rotational diffusion, the following can be observed. 
For $\D=0$, $\chi_3^{(\a)}(\om)$ for the two models are very similar, cf. Fig.\ref{Plot2} and Figs.3,4 of ref.\cite{Dejardin00}.
For finite $\D$, however, the third-order response for the ADWP-model shows a characteristic temperature dependence, that is absent in the model of rotational Brownian motion.

In Fig.\ref{Plot2}, the real and the imaginary part of the 3$\om$-component $\chi_3^{(3)}(\om)$ are plotted 
versus $\om\t$ for different values of the asymmetry $\D$ and various temperatures.
It is evident that the sign of both functions change as a function of frequency. 
Furthermore, the shapes of $\rm Im(\chi_3^{(3)}(\om))$ differ significantly from Lorentzians.
As mentioned above, for $\D=0$, $\chi_3^{(3)}(\om)$ does not depend on temperature. 

The static nonlinear susceptibilites are determined by the limiting values of the spectral dfunctions, $S_3^{(1)}(0)=(3\d^2-1)$ and $S_3^{(3)}(0)=(3\d^2-1)/3$, and thus are given by:
\be\label{Chi3.0.ADWP}
\chi_3^{(3)}(0)={M^4\over60}\b^3\left(3\d^2-1\right)\left(1-\d^2\right)
\quad;\quad
\chi_3^{(1)}(0)=3\chi_3^{(3)}(0)
\ee
It should be mentioned, that $\chi_3^{(\a)}(0)$ is determined by the fourth-order cumulant,
$\k_4(M)=\lg M^4\rg-4\lg M\rg\lg M^3\rg-3\lg M^2\rg^2+12\lg M\rg^2\lg M^2\rg-6\lg M\rg^4
=2M^4\left(3\d^2-1\right)\left(1-\d^2\right)$.
For finite $\D$, the low-frequency limit $\chi_3^{(\a)}(0)$ vanishes at a temperature $T_0$, at which $S_3^{(\a)}(0)=0$, 
\[
T_0=\D/\ln{[(\sqrt{3}+1)/(\sqrt{3}-1)]}\simeq\D/1.317.
\]
For large frequencies, one always has $\chi_3^{(\a)}(\infty)=0$.

\begin{figure}[h!]
\centering
\includegraphics[width=8.0cm]{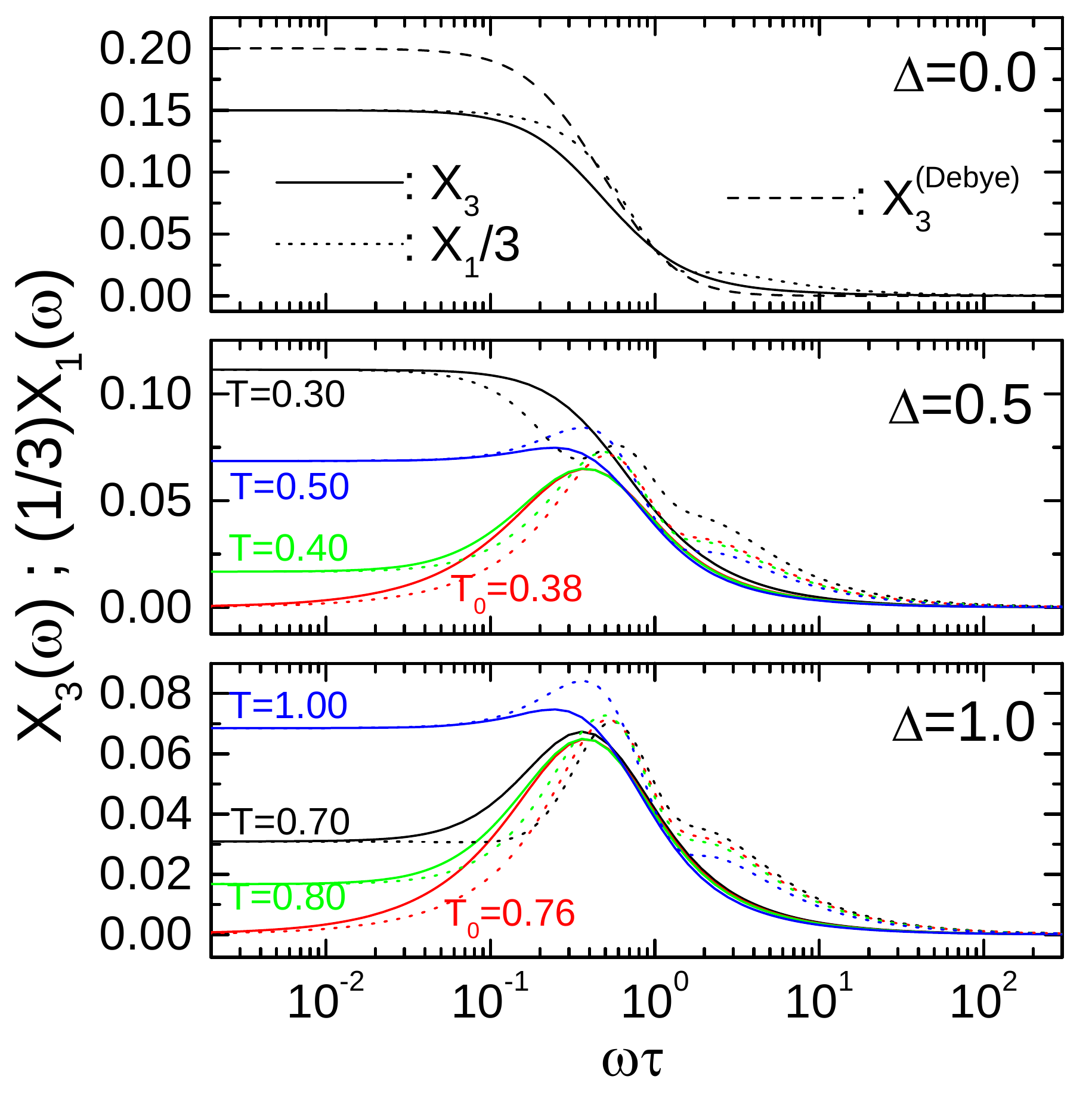}
\vspace{-0.5cm}
\caption{$X_\a(\om,T)$ for various values of the asymmetry and different temperatures.
In the uppermost panel, $X_3^{(\rm Debye)}(\om)$\cite{Dejardin00} is shown for comparison (dashed line). }
\label{Plot3}
\end{figure}

Instead of discussing $\chi_3^{(\a)}(\om)$ further, in the following I will consider $X_\a(\om,T)$ according to eq.(\ref{Xalfa.Def}).
This quantity is given by, cf. eq.(\ref{Chi1.ADWP}) and eq.(\ref{Chi3.ADWP}):
\be\label{Xalfa.ADWP}
X_\a(\om,T)={9\over20}{\left|S_3^{(\a)}(\om\t)\right|\over\left(1-\d^2\right)}
\ee
The limiting values for small and large frequencies are determined by the corresponding limits of 
$S_3^{(\a)}(\om\t)$ and thus, one has for example
$X_3(0,T)=(3/20)(\left|3\d^2-1\right|/\left(1-\d^2\right))$.
It is evident, that $X_\a(\om,T)$ will have a peak-like structure for $T\simeq T_0$.
As is shown in Fig.\ref{Plot3}, for other temperatures one has 'trivial' behavior, i.e. a continuous decay from the low-frequency limit to $X_\a(\om,T)=0$ at high frequencies.
One can see, that the behavior of the $1\om$-component and the $3\om$-component is very similar.

\begin{figure}[h!]
\centering
\includegraphics[width=8.0cm]{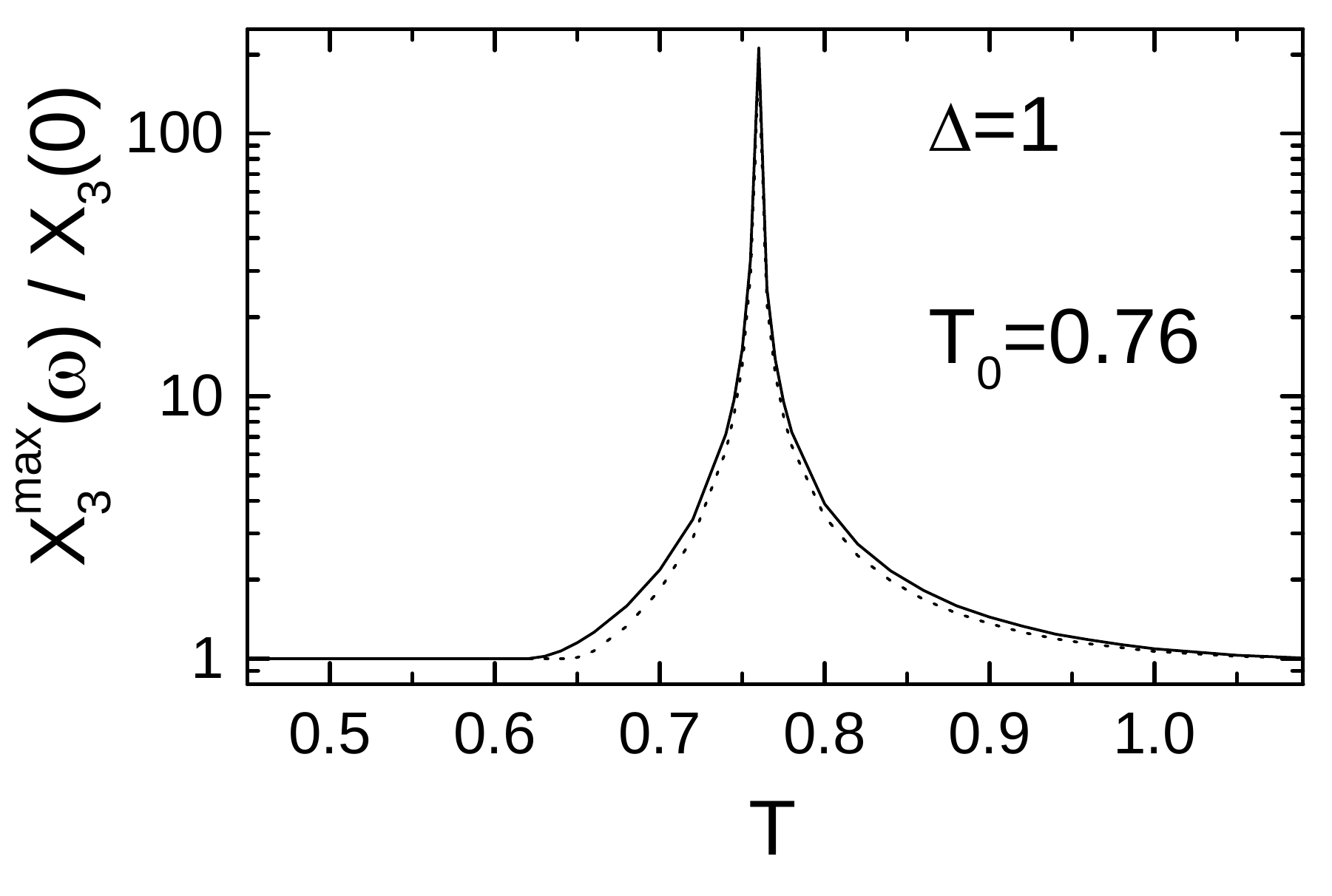}
\vspace{-0.5cm}
\caption{$X_3^{\rm max}(\om)/X_3(0)$ versus temperatures for $\D=1$.
The dotted line is the same with assumption of a Gaussian distribution of $\D$ with mean $\overline\D=1$ and variance $\s_\D=10$.}
\label{Plot4}
\end{figure}

In order to further quantify the behavior of $X_\a(\om,T)$ with regard to a 'hump'-like structure, in Fig.\ref{Plot4}, the ratio $X_3^{\rm max}(\om)/X_3(0)$ is plotted versus temperature.
For $T\ll T_0$ and also for $T\gg T_0$ trivial behavior is observed and only in the region of $T\sim T_0$ a hump develops. 
This hump, however, has nothing to do with glassy correlations but is solely a consequence of the temperature dependence of the fluctuations of the dipole moments.

Finally, it is to be mentioned that the above results hardly change if one considers distributions of the hopping rate $W$ and/or the asymmetry.
In particular, the temperature-dependent change in the shape of $X_\a(\om)$ is practically unaltered.
This is exemplified in Fig.\ref{Plot4}, where the dotted line represents $X_3^{\rm max}(\om)/X_3(0)$ for the case of a broad Gaussian distribution of $\D$.
The reason for this is simply the steepness of the root of $S_3^{(\a)}(0)=0$, meaning that the overall behavior is determined by the mean value of $\D$.
Thus, if one considers a system with a distribution of asymmetries that is centered at $\D=0$, one will observe trivial behavior of $X_\a(\om)$ at all temperatures.
Ladieu et al. use the ADWP-model with finite $\D$ and some further assumptions to fit the experimental data on supercooled liquids\cite{Ladieu12}.
\section*{IV. Trap models}
In this section, I will discuss $X_\a(\om,T)$ for the trap model with a Gaussian density of states, which, 
as already mentioned in the Introduction, shows some features of glassy relaxation.
It is defined by the ME for $G(\e,t+t_0|\e_0,t_0)=G(\e,t|\e_0,0)\equiv G(\e,t|\e_0)$, in a continuous form written as:
\be\label{ME.G}
{\dot G}(\e,t|\e_0)= -\k(\e)G(\e,t|\e_0)+\rho(\e)\!\int\!d\e'\k(\e')G(\e',t|\e_0) 
\ee
In eq.(\ref{ME.G}), the escape rate is given by
\be\label{k.T}
\k(\e)=\k_\infty e^{\b\e}
\ee
with the attempt rate $\k_\infty$.
Furthermore, I solely consider the model with a Gaussian DOS
\be\label{DOS.Gauss}
\rho(\e)\!=\!{1\over\sqrt{2\pi}\s}e^{-\e^2/(2\s^2)}
\ee
with $\s=1$.
From eq.(\ref{ME.G}), the equilibrium populations at a given temperature $T$ (measured in units of $\s$) are found to be Gaussian 
$p^{\rm eq}(\e)=\lim_{t\to\infty}G(\e,t|\e_0)={1\over\sqrt{2\pi}\s}e^{-(\e-{\bar\e})^2/(2\s^2)}$
with ${\bar\e}=-\b \s^2$.

In order to calculate the response, one further has to quantify the dependence of the dynamical variable on the trap energy $\e$. 
The choice of this dependence represents a further assumption of the calculation and has a strong impact on the results for the cubic response, as will be discussed below.
In order to clarify this issue, consider the linear response for the specific choice of eq.(\ref{Wkl.HX}) for the field-dependence of the transition rates.
Using eqns.(\ref{F.expect}), (\ref{Lkl.def}) and (\ref{Chi1}), one obtains the relation between the linear response and the equilibrium auto-correlation function $C_M(t)=\lg M(t)M(0)\rg$,
$R^{(1)}_M(t)=-\b(\g+\mu)[dC_M(t)/dt]$, if the system is in thermal equilibrium \cite{G54}.
In the frequency-domain, this yields eq.(\ref{Chi1.Trap}) in Appendix B, if the average over the possible realizations of the variables is performed with the following assumption:
\be\label{MkMl.mit}
\lg M(\e)\rg=0
\quad\mbox{and}\quad
\lg M(\e)M(\e_0)\rg=\d(\e-\e_0)\lg M(\e)^2\rg
\ee
In the calculation of the third-order response, the fourth moments of the variable are important.
For the corresponding averages I will assume a Gaussian factorization property for simplicity:
\Be\label{Mh4.mit.Gauss}
\lg M(\e_1)M(\e_2)M(\e_3)M(\e_4)\rg
&&\hspace{-0.6cm}=
\d(\e_1-\e_2)\d(\e_3-\e_4)\lg M(\e_1)^2\rg\lg M(\e_3)^2\rg
\nonumber\\
&&\hspace{-0.6cm}+\;
\d(\e_1-\e_3)\d(\e_2-\e_4)\lg M(\e_1)^2\rg\lg M(\e_2)^2\rg
\\
&&\hspace{-0.6cm}+\;
\d(\e_1-\e_4)\d(\e_2-\e_3)\lg M(\e_1)^2\rg\lg M(\e_2)^2\rg
\nonumber
\Ee
In the calculation of the response, the field-dependence of the transition rates has to be fixed additionally.
I use eq.(\ref{Wkl.HX}) with arbitrary values for $\g$ and $\mu$. 
From the physics of the model one might argue that $\mu=1$ and $\g=0$ is an appropriate choice because it is meaningful to assume that the activation energy of the escape is biased by the field (according to 
$\e\to\e-M(\e)\cdot H)$.
However, it is not clear that this simple argument holds in out-of-equilibrium situations and for strong fields.
Using the assumptions made, one can compute the response according to the expressions given in Appendix A.
The calculation is outlined in Appendix B and here only the results will be discussed.

In the explicit choice of the variable, I follow Fielding and Sollich\cite{FS02} and use a set of variables with an Arrhenius-like dependence on the trap energies:
\be\label{Mh2.mit.n}
\lg M(\e)^2\rg=e^{-n\b\e}
\ee
with variable $n$ and where the static value of $M^2$ has been set to unity.
For $n=0$, one has temperature-independent variables as in case of Brownian rotational diffusion.

The most important consequence of the specific choice, eq.(\ref{Mh2.mit.n}), is that it does not affect the spectral shape of the linear response.
The only quantities that strongly depend on the choice of $n$ are the static susceptibility and the
the temperature dependence of the relaxation time.
This is because one can write:
\[
\int\!d\e p(\e)^{\rm eq}e^{-n\b\e}{\k(\e)\over\k(\e)-i\om}
=e^{{n(n+2)\over2}\b^2\s^2}
\int\!d\e p(\e)^{\rm eq}{\k(\e)\over\k(\e)-i\om_n}
\]
with
\be\label{om.n.def}
\om_n=\om e^{n\b^2\s^2}
\ee
Thus, the susceptibility is given by:
\be\label{Chi1.n.Trap}
\chi_1(\om)=\b(\g+\mu)\int\!d\e p(\e)^{\rm eq}e^{-n\b\e}{\k(\e)\over\k(\e)-i\om}
=\D\chi_1\int\!d\e p(\e)^{\rm eq}{\k(\e)\over\k(\e)-i\om_n}
\ee
The static susceptiblity, i.e. the amplitude, $\D\chi_1$, strongly depends on the choice of $n$ and reads as:
\be\label{DChi1.n.Trap}
\D\chi_1=(\g+\mu)\b \overline{\lg M^2\rg}_T
=(\g+\mu)\b e^{{n(n+2)\over2}\b^2\s^2}
\ee
Here, the second moment $\overline{\lg M^2\rg}_T$ is related to the low-frequency limit of $\chi_1(\om)$, 
$\overline{\lg M^2\rg}_T=\int\!d\e\lg M(\e)^2\rg p(\e)^{\rm eq}$.
Note that $\D\chi_1$ is temperature independent only for $n=0$ and for $n=-2$.

\begin{figure}[h!]
\centering
\includegraphics[width=8.0cm]{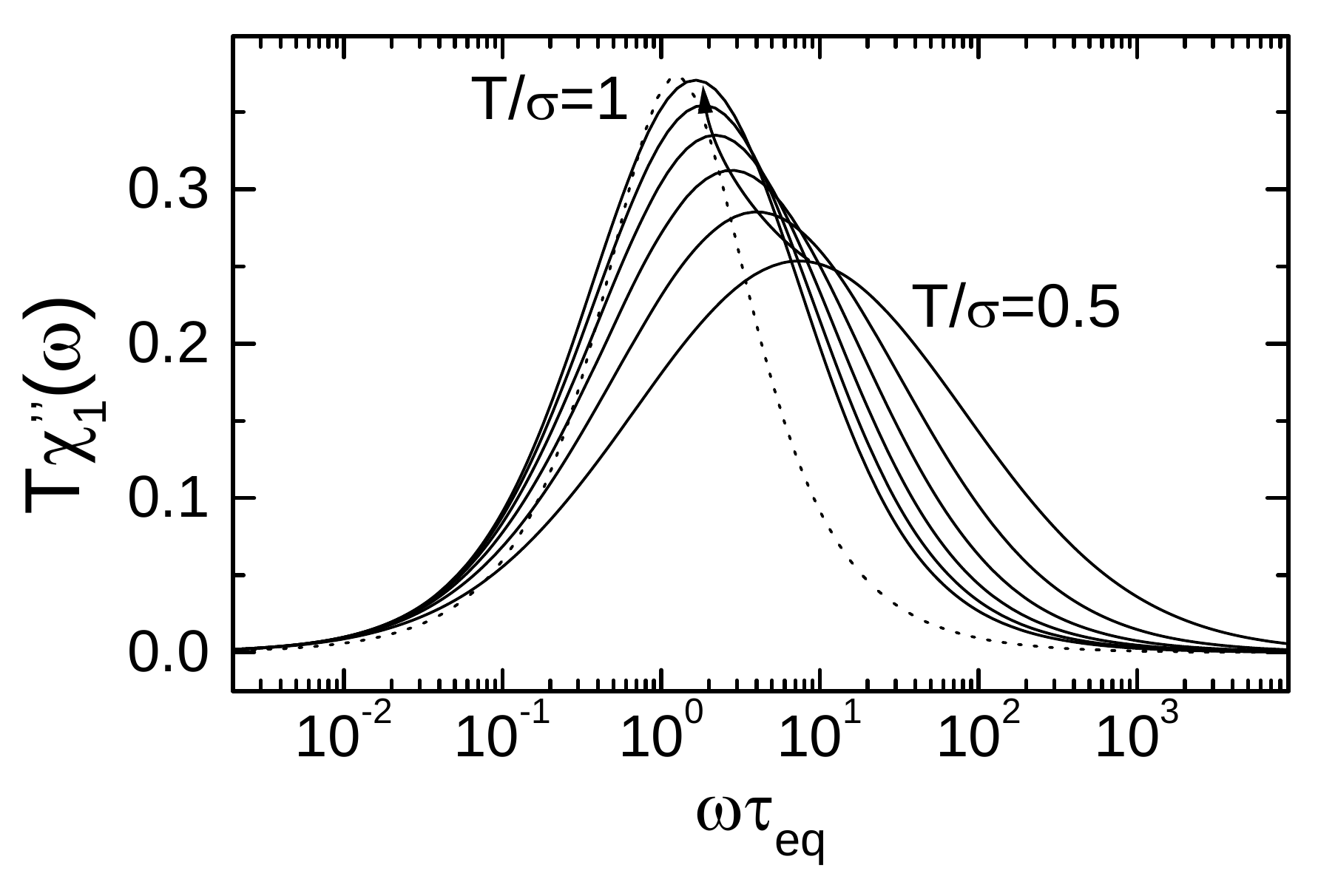}
\vspace{-0.5cm}
\caption{Imaginary part of $T\chi_1(\om)$, $T\chi_1''(\om)$, for $n=0$ and various temperatures ($T/\s$=0.5, 0.6, 0.7, 0.8, 0.9, 1.0 as indicated by the arrow). 
The dotted line represents a Lorentzian.}
\label{Plot5}
\end{figure}

In Fig.\ref{Plot5}, the imaginary part of $\chi_1(\om)$ is shown for $n=0$ and various temperatures.
The frequencies are scaled to the relaxation time of $C_M(t)$ for $n=0$, 
$\t_{\rm eq}=\int_0^\infty\!\!dtC_M(t)=\k_\infty^{-1}e^{{3\over2}\b^2\s^2}$, cf. ref.\cite{G64}.
It is obvious that $\chi_1''(\om)$ broadens as temperature is decreased and thus time-temperature-superposition is not obeyed. 
It is stressed again, that $\chi_1(\om)$ is basically independent of the choice of $n$.

Next, the behavior of the cubic response and its dependence on the model parameters will be discussed.
Using the limiting values of the cubic response functions given in Appendix B for small and high frequencies, one finds the following limits for $\chi_3^{(\a)}(\om)$:
\be\label{Chi3n.Limit.Trap}
\chi_3^{(3)}(0)={1\over8}\b^3(\g+\mu)^3(\xi_2-\xi_1)
\quad;\quad 
\chi_3^{(1)}(0)=3\chi_3^{(3)}(0)
\quad\mbox{and}\quad 
\chi_3^{(\a)}(\infty)=0
\ee
Here, I defined the averages $\xi_1=\overline{\lg M^2\rg}_\infty\overline{\lg M^2\rg}_T$ and
$\xi_2=\overline{\lg M^2\rg^2}_T$, which for the Gaussian trap model are given by:
\be\label{ksi.1.2.Trap}
\xi_1=e^{n(n+1)\b^2\s^2}
\quad;\quad
\xi_2=e^{2n(n+1)\b^2\s^2}
\ee
With these quantities, one finds for the low-frequency limit of $X_3$:
\be\label{X3.0.Trap}
X_3(0,T)={1\over8}(\g+\mu){|\xi_2-\xi_1|\over\left(\overline{\lg M^2\rg}_T\right)^2}
\ee
and similarly for $X_1(0,T)$.
It is thus clear that these low-frequency limits do strongly depend on the variable, i.e. on $n$.
Therefore, one can expect to find trivial or hump-like behavior of $X_\a(\om,T)$, $\a=1, 3$, depending on this choice.

\begin{figure}[h!]
\centering
\includegraphics[width=8.0cm]{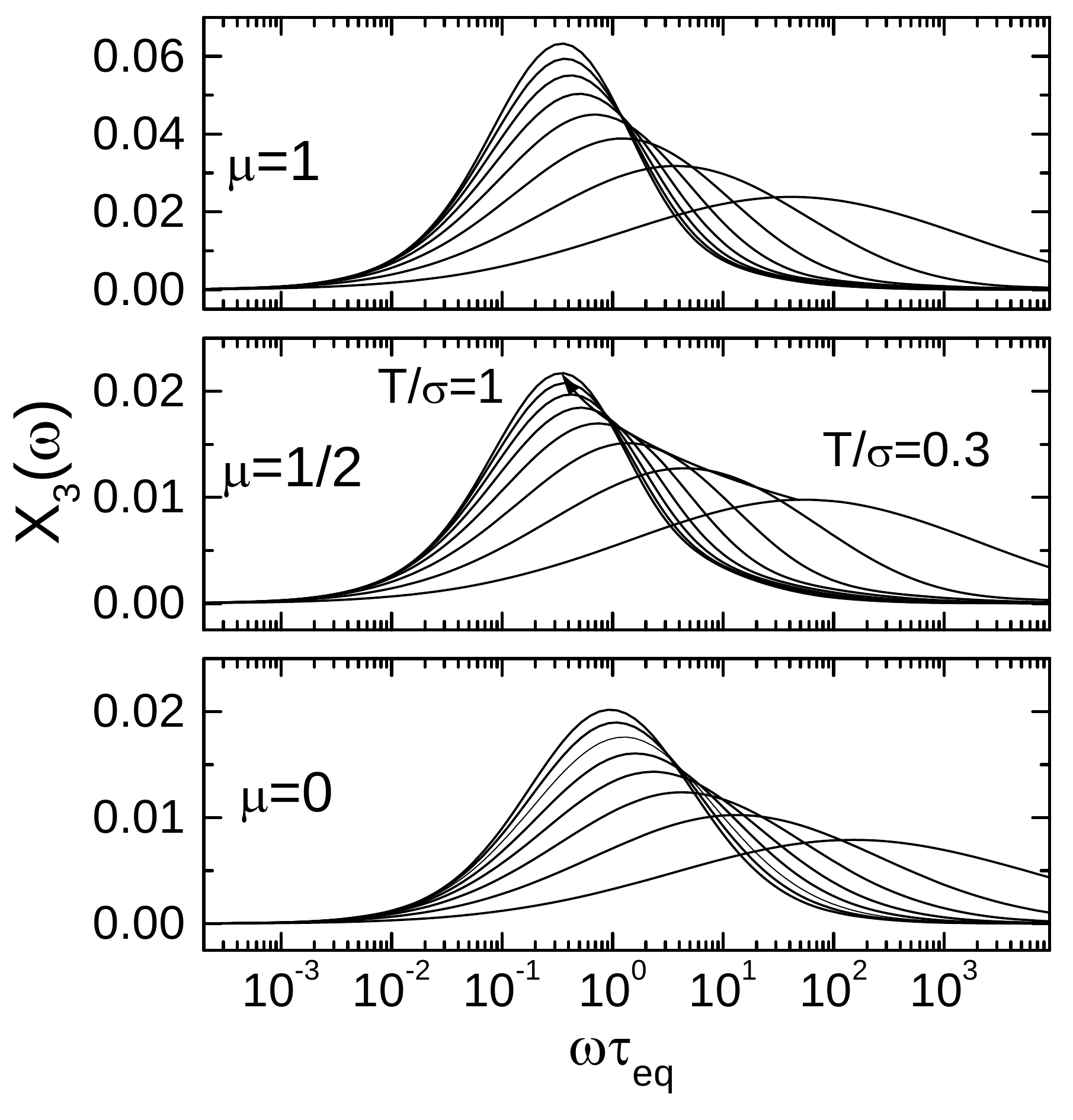}
\vspace{-0.5cm}
\caption{$X_3(\om,T)$ for $n=0$ and various values of $\mu$ for $\g=1-\mu$ and different temperatures
($T/\s=0.3, 0.4, 0.5, 0.6, 0.7, 0.8, 0.9, 1$) in the order indicated by the arrow.}
\label{Plot6}
\end{figure}

In Fig.\ref{Plot6} $X_3(\om,T)$ is shown for $n=0$ and various values of $\mu$. Here, it is assumed that 
$\g+\mu=1$.
The main difference between the various choices for $\mu$ is the overall amplitude. 
Additionally, it is clear that $X_3(\om,T)$ exhibits a hump in all cases.
However, in contrast to the results obtained on supercooled liquids, the maximum value of $X_3$ increases as a function of temperature.
This increase is somewhat stronger for $\mu=1$ than it is for other values of $\mu$.

\begin{figure}[h!]
\centering
\includegraphics[width=8.0cm]{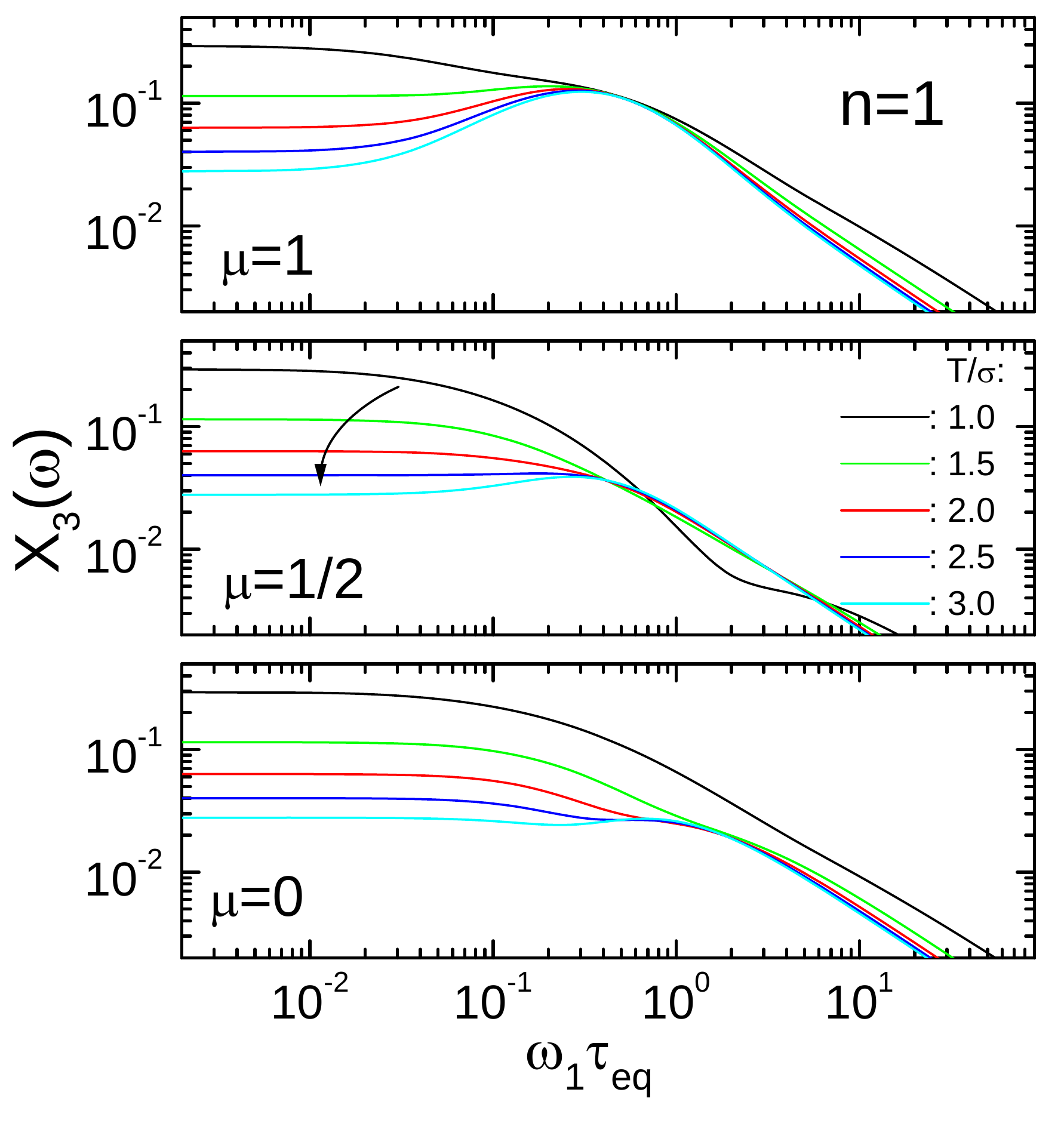}
\hspace{0.5cm}
\includegraphics[width=8.0cm]{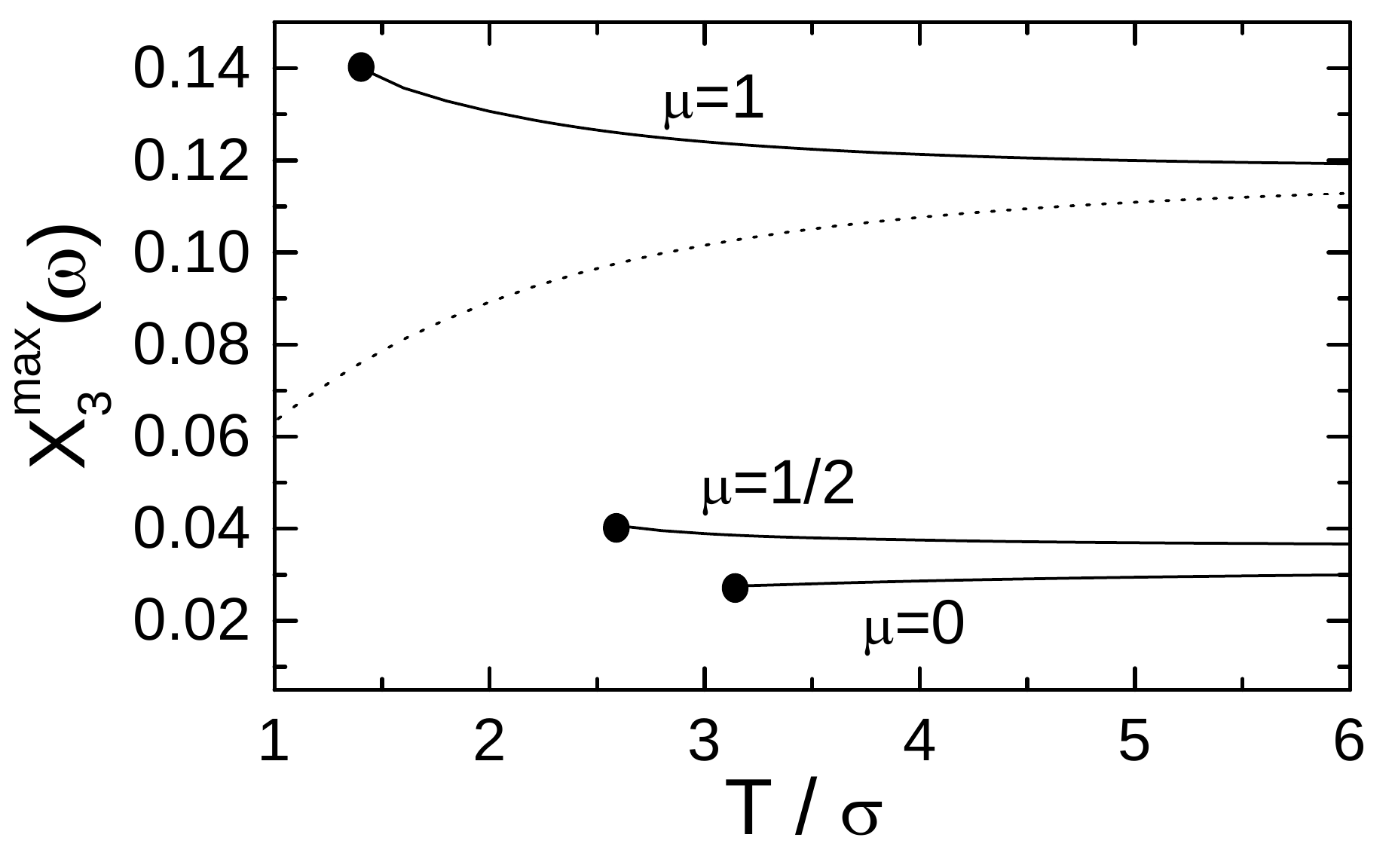}
\vspace{-0.5cm}
\caption{{\bf a}: (left) $X_3(\om,T)$ for $n=1$ and different temperatures ($T/\s=1, 1.5, 2, 2.5, 3$). 
The arrow indicates increasing temperature.
{\bf b}: (right) $X_3^{\rm max}(\om,T)$ as a function of temperature for $n=1$. 
The curves are shown for temperatures higher than the onset temperature, below which trivial behavior is observed, i.e. $\om_{\rm max}=0$.
The dotted line is the result for $n=0$, $\mu=1$.}
\label{Plot7}
\end{figure}

Next, I will consider values for $n$ different from zero, meaning that the dynamical variable that couples to field shows an explicit dependence on the trap energies.
In Fig.\ref{Plot7}a, $X_3(\om)$ is plotted versus frequency for $n=1$ and the same values for $\mu$ as in 
Fig.\ref{Plot6}.
It is observed that a hump is found at high temperatures, whereas trivial behavior is observed at low temperatures. 
The temperature, at which a visible peak is observed depends on the value of $\mu$, i.e. on the way, the field couples to the transition rates.
This is shown in Fig.\ref{Plot7}b, where the maximum value of $X_3(\om)$ is plotted versus temperature for temperatures higher than the onset temperature, which is defined by the first appearance of a hump in $X_3(\om)$
indicated by the dots in Fig.\ref{Plot7}b.
In the temperature range of a hump-like shape of $X_3(\om,T)$ its maximum, $X_3^{\rm max}(\om,T)$, appears to be almost independent of temperature.
A similar behavior is found for other positive values of $n$.

From these model calculations it becomes apparent that the existence of a hump depends on the value of 
$X_\a(0)$, the value of the maximum of $X_\a(\om)$, and in particular their ratio.
Thus, the low-frequency limit plays an important role in determining the overall shape of $X_\a(\om)$.

\begin{figure}[h!]
\centering
\includegraphics[width=8.0cm]{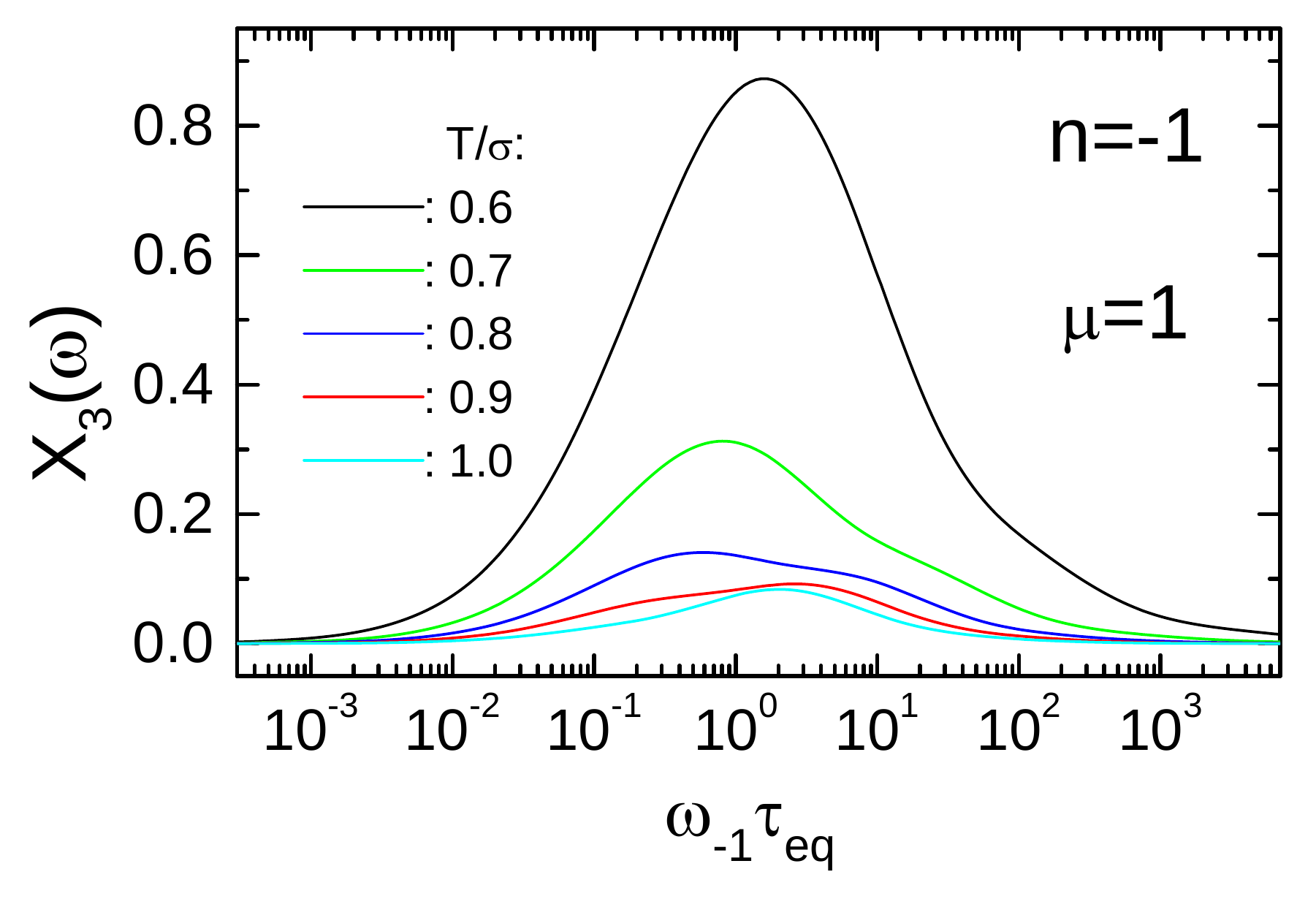}
\hspace{0.5cm}
\includegraphics[width=8.0cm]{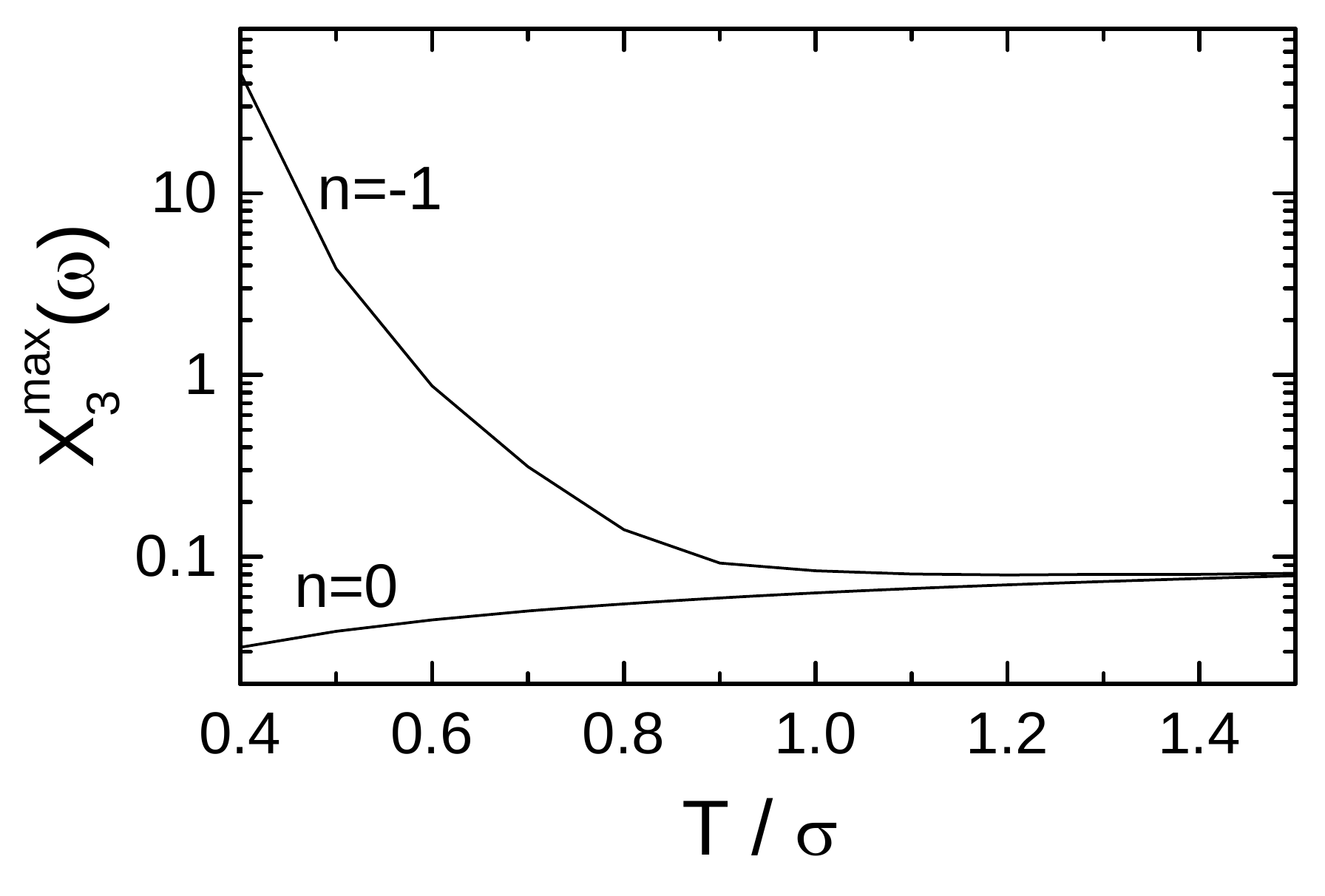}
\vspace{-0.5cm}
\caption{{\bf a}: (left) $X_3(\om,T)$ for $n=-1$, $\mu=1$ and different temperatures 
($T/\s=0.6, 0.7, 0.8, 0.9, 1$ from top to bottom).
{\bf b}: (right) $X_3^{\rm max}(\om,T)$ as a function of temperature for $n=-1$ and $n=0$ ($\mu=1$).}
\label{Plot8}
\end{figure}

These considerations can be further substantiated by considering the special value of $n=-1$, because in this case one has $\xi_1=\xi_2=0$ and therefore $X_\a(0)=0$, cf. eq.(\ref{ksi.1.2.Trap}).
This means, a hump will be observed in this case, as is confirmed in Fig.\ref{Plot8}a, where $X_3(\om)$ is plotted as a function of frequency for $\mu=1$. 
For other values of $\mu$, the results are very similar.
On first sight, the behavior of $X_3(\om)$ is very similar to that for $n=0$, cf. Fig.\ref{Plot6}.
However, the maximum for $n=-1$, $X_3^{\rm max}(\om,T)$, is a decreasing function of temperature as opposed to the case of $n=0$, cf. Fig.\ref{Plot8}b.

At this point, however, it has to be noted that the case $n=-1$ is somewhat special as the mean relaxation time of the linear response, 
$\lg\t\rg=\int\!d\e p(\e)^{\rm eq}(e^{-\b\e}/\k(\e))=\k_\infty^{-1}$, is basically temperature-independent.
Thus, although the shapes of $\chi_1(\om)$ are identical for $n=0$ and $n=-1$ at a given temperature, the mean relaxation time for $n=-1$ does not change with temperature.
This shows that it is not straightforward to compare linear and nonlinear response functions.
\section*{V. Conclusions}
The theory of nonlinear response functions for a system obeying a master equation has been formulated in close analogy to quantum mechanical nonlinear response theory.
Time-dependent perturbation theory is used in order to compute the elements of the propagator (the Green's function or conditional probability) in the desired order of the amplitude of the applied external field.
Expressions for the response functions up to third order are given in terms of the solution of the field-free master equation for systems with arbitrary initial conditions and also for non-stationary Markov processes.
In the actual model calculations, however, only stationary systems are considered that were in thermal equilibrium prior to the application of the field. 
The treatment of aging systems or other non-equilibrium situations are beyond the scope of the present paper.

For the model of dipole reorientations in an asymmetric double well potential (ADWP-model), the spectral shape of the modulus of the frequency-dependent cubic response, $X_3(\om,T)$, shows a specific temperature dependence which strongly depends on the value of the static susceptibility, $X_3(0,T)$.
At a temperature $T_0$, which is determined by the value of the asymmetry of the potential, $X_3(0,T_0)$ vanishes.
For a narrow temperature range in the vicinity of $T_0$ a peak is observed in the modulus.
For temperatures sufficiently different from $T_0$ a monotonous decay from $X_3(0)\neq0$ to $X_3(\infty)=0$ is found.
This 'trivial' behavior is basically the same as for the model of rotational Brownian motion\cite{Dejardin00} and is at variance with experimental results obtained for supercooled glycerol\cite{CrausteThibierge10, Brun11}.
It was attributed to trivial dipole reorientations that occur independent of glassy correlations.
These correlations should give rise to a peaked behavior, i.e. the existence of a hump in $X_3(\om)$.
If one intends to utilize the ADWP model for the dipole reorientations in supercooled liquids, it is natural to assume distributions of relaxation rates and of asymmetries.
However, as shown in Section III, such a distribution hardly affects the spectral shape of $X_3(\om)$ (apart from the fact that a distribution of relaxation times gives rise to a broadening).

If a trap model with a Gaussian distribution of trap energies is considered, a more complex dependence of $X_3(\om)$ on the parameters used in the calculations is observed.
In particular, the dependence of the dynamical variables that couple to the external field on the trap energies, $M(\e)$, has to be fixed.
I restricted the calculations to variables that obey Gaussian statistics and depend on the trap energy in an exponential way, $M(\e)=e^{-n\b\e}$, cf. eq.(\ref{Mh2.mit.n}).
This choice is particularly useful when discussing the properties of nonlinear response functions and their relation to the linear response.
This is because the exponential dependence on the trap energies has the interesting property that the spectral shape of the linear susceptibility is the same for all values of the parameter $n$.
Only the amplitude ($\D\chi_1$) and the temperature dependence of the relaxation time strongly depend on its specific value.
If the nonlinear response is considered, it is however found that the temperature-dependent spectral shape of $X_3(\om)$ strongly depends on the value of $n$.
In particular, one can find a peak or 'trivial' behavior depending on both, the value of $n$ and the temperature.
Similar to the situation in the ADWP-model, the existence of a peak is related to the value of the static susceptibility.
In case of the occurrence of a hump, the temperature dependence of the peak maximum, $X_3^{\rm max}(\om)$, can increase ($n=0$) or decrease ($n=-1$) with increasing temperature.
These results indicate that it is difficult to compare the linear and nonlinear susceptibilities.
It is left for future work to investigate the behavior of the nonlinear response in the trap model for other dynamical variables and also for non-equilibrium situations. 

In the experimental determination of $X_3(\om,T)$ in supercooled liquids\cite{CrausteThibierge10, Brun11}, the decrease of $X_3^{\rm max}(\om)$ with increasing temperature has been used to extract the number of correlated molecules, $N_{\rm corr}$, which is a 'real space property' of the dynamical heterogenities in glasses.
Due to the mean-field nature of both models considered in the present paper, none of the results presented have any connection to real space.
Therefore, a direct comparison to experimental data is not possible.
However, the model calculations substantiate the fact observed earlier already\cite{Brun11b} that the existence of a peak in $X_3(\om)$ does not have to be related to glassy correlations in some sense.

In conclusion, I have formulated a theory of nonlinear response for systems described by Markov processes and have presented the results of calculations for simple stochastic models.
The most important result is that the spectral shape of the nonlinear (cubic) response can vary considerably depending on the model considered.
The occurrence of a peak in the modulus of the third-order susceptibility cannot generally be attributed to glassy correlations.
Of course, this does not mean that glassy correlations do not give rise to a hump but its mere existence cannot be taken as a signature of such correlations.
Because the models considered in the present paper are of a mean-field nature, it is impossible to connect the results to a length scale of any kind.
Due to the growing interest in nonlinear responses in complex systems, calculations of the kind presented in the present paper should be performed for a variety of different models in order to gain a deeper understanding of the general features governing the shape and the temperature dependence of the corresponding 
susceptibilities.
\section*{Acknowledgment}
I thank Roland B\"ohmer, Gerald Hinze, Francois Ladieu and Jeppe Dyre for fruitful discussions and 
Roland B\"ohmer for helpful comments on the mansucript.
\newpage
\begin{appendix}
\section*{Appendix A: Calculation of nonlinear response functions}
\setcounter{equation}{0}
\renewcommand{\theequation}{A.\arabic{equation}}
In this appendix the calculation of the response for a system obeying the ME, eq.(\ref{ME.t.abh}), using time-dependent perturbation theory is described.
Using eq.(\ref{We.kl.t}) for the master-operator, the ME in a matrix notation reads:
\be\label{ME.matrix}
\partial_t\G(t,t_0)={\cal W}(t)\G(t,t_0)
\ee
Here, the propagator has matrix-elements $\G(t,t_0)_{kl}=G_{kl}(t,t_0)$.
The solution of the ME in the absence of an external field can be written in the form:
\be\label{G.LsgME.allg}
\G(t,t_0)={\cal T}\exp{\left(\int_{t_0}^td\t{\cal W}(\t)\right)}\G(t_0,t_0)
\ee
where ${\cal T}$ denotes the time-ordering operator and $\G(t_0,t_0)_{kl}=\d_{kl}$.
In the presence of the field the transition rates are given by eq.(\ref{W.H.kl.Taylor}) and the corresponding master-operator accordingly reads as ${\cal W}^{(H)}(t)_{kl}=W^{(H)}_{kl}(t)-\d_{kl}\sum_nW^{(H)}_{nl}(t)$.
The ME is written as $\partial_t\G^{(H)}(t,t_0)={\cal W}^{(H)}(t)\G^{(H)}(t,t_0)$.

In order to calculate the response of the system to an external field applied at time $t\!=\!t_0$ and measured by an observable $F(t)$, $\lg F(t)\rg_{(H)}=\sum_{kl}F_kG^{(H)}_{kl}(t,t_0)p_k(t_0)$ as given in eq.(\ref{F.expect}), time-dependent perturbation theory is used to express the propagator as a series of the form
$\G^{(H)}(t,t_0)=\G(t,t_0)+\sum_{n=1}^\infty\G^{(n)}(t,t_0)$, where $\G(t,t_0)$ denotes the propagator in the field-free case.
In order to perform the calculation, one proceeds in the following way.
Starting from the Dyson-like equation
\be\label{G.Dyson}
\G^{(H)}(t,t_0)=\G(t,t_0)+\int_{t_0}^t\!dt'\G(t,t'){\cal V}(t')\G^{(H)}(t',t_0)
\ee
one obtains, using eq.(\ref{W.H.Vn}), for the lowest order terms:
\Be\label{G.Reihe}
\G^{(1)}(t,t_0)
&&\hspace{-0.6cm}=
\int_{t_0}^t\!dt'\G(t,t'){\cal V}^{(1)}(t')\G(t',t_0)
\nonumber\\
\G^{(2)}(t,t_0)
&&\hspace{-0.6cm}=
\int_{t_0}^t\!dt'\G(t,t'){\cal V}^{(2)}(t')\G(t',t_0)
+\int_{t_0}^t\!dt'\G(t,t'){\cal V}^{(1)}(t')\G^{(1)}(t',t_0)
\nonumber\\
\G^{(3)}(t,t_0)
&&\hspace{-0.6cm}=
\int_{t_0}^t\!dt'\G(t,t'){\cal V}^{(3)}(t')\G(t',t_0)
+\int_{t_0}^t\!dt'\G(t,t'){\cal V}^{(1)}(t')\G^{(2)}(t',t_0)
\\
&&\hspace{4.82cm}
+\int_{t_0}^t\!dt'\G(t,t'){\cal V}^{(2)}(t')\G^{(1)}(t',t_0)
\nonumber
\Ee
In the next step, one uses the expression for the matrix elements of $\G^{(n)}(t,t_0)$, denoted by
$G^{(n)}_{kl}(t,t_0)$, in eq.(\ref{F.expect}) in order to compute the nth-order response,
$\chi^{(n)}_F(t,t_0)$.

With the definition
\be\label{Lkl.def}
L_{kj}^{(\eta)}(t_2,t_1)=\sum_m\left[G_{km}(t_2,t_1)-G_{kj}(t_2,t_1)\right]W_{mj}^{(\eta)}(t_1)
\ee
where $W_{mj}^{(n)}(t_1)$ is given in eq.(\ref{W.H.kl.Taylor}), one obtains in a straightforward calculation for the linear response:
\be\label{Chi1}
\chi^{(1)}_F(t,t_0)=\int_{t_0}^t\!dt_1H(t_1)R^{(1)}_F(t,t_1)
\quad\mbox{with}\quad
R^{(1)}_F(t,t_1)=\b\sum_{k,l}F_kL_{kl}^{(1)}(t,t_1)p_l(t_1)
\ee
From the structure of this expression it is evident that $R$ denotes the usual response to a short field kick.

The second-order response is found to consist of two terms:
\be\label{Chi2}
\chi^{(2)}_F(t,t_0)=\chi^{(2;1)}_F(t,t_0)+\chi^{(2;2)}_F(t,t_0)
\ee
with 
\Be\label{R2.def}
\chi^{(2;1)}_F(t,t_0)
&&\hspace{-0.6cm}=
\int_{t_0}^t\!dt_1H(t_1)\int_{t_0}^{t_1}\!dt_2H(t_2)R^{(2;1)}_F(t,t_1,t_2)
\nonumber\\
&&\hspace{-0.6cm}
R^{(2;1)}_F(t,t_1,t_2)=
\b^2\sum_{k,l,m}F_kL_{km}^{(1)}(t,t_1)L_{ml}^{(1)}(t_1,t_2)p_l(t_2)
\nonumber\\
\chi^{(2;2)}_F(t,t_0)
&&\hspace{-0.6cm}=
{1\over2}\int_{t_0}^t\!dt_1H(t_1)^2R^{(2;2)}_F(t,t_1)
\\
&&\hspace{-0.6cm}
R^{(2;2)}_F(t,t_1)=
\b^2\sum_{k,l}F_kL_{kl}^{(2)}(t,t_1)p_l(t_1)
\nonumber
\Ee
This second order response is expected to be of little relevance in most cases as it vanishes in isotropic systems.
More interesting is the third-order response because usually this is the lowest-order nonlinear contribution to the response of the system. 
As can be expected from Fig.\ref{Plot1}, it has the form:
\be\label{Chi3}
\chi^{(3)}_F(t,t_0)=\chi^{(3;1)}_F(t,t_0)+\chi^{(3;2)}_F(t,t_0)+\chi^{(3;3)}_F(t,t_0)
\ee
and the individual terms are given by:
\Be\label{R31.def}
\chi^{(3;1)}_F(t,t_0)
&&\hspace{-0.6cm}=
\int_{t_0}^t\!dt_1H(t_1)\int_{t_0}^{t_1}\!dt_2H(t_2)\int_{t_0}^{t_2}\!dt_3H(t_3)
R^{(3;1)}_F(t,t_1,t_2,t_3)
\\
&&\hspace{-0.6cm}
R^{(3;1)}_F(t,t_1,t_2)=
\b^3\!\sum_{k,l,m,n}F_kL_{km}^{(1)}(t,t_1)L_{mn}^{(1)}(t_1,t_2)L_{nl}^{(1)}(t_2,t_3)p_l(t_3)
\nonumber
\Ee
originating from the linear perturbation. This term also is found in the response theory for a Fokker-Planck equation.
The cross-terms between the first- and second-order perturbations are:
\Be\label{R32.def}
\chi^{(3;2)}_F(t,t_0)
&&\hspace{-0.6cm}=
\chi^{(3;2A)}_F(t,t_0)+\chi^{(3;2B)}_F(t,t_0)
\nonumber\\
&&\hspace{-0.6cm}
\chi^{(3;2A)}_F(t,t_0)=
{1\over2}\int_{t_0}^t\!dt_1H(t_1)\int_{t_0}^{t_1}\!dt_2H(t_2)^2R^{(3;2A)}_F(t,t_1,t_2)
\nonumber\\
&&\hspace{1.6cm}
R^{(3;2A)}_F(t,t_1,t_2)=
\b^3\sum_{k,l,m}F_kL_{km}^{(1)}(t,t_1)L_{ml}^{(2)}(t_1,t_2)p_l(t_2)
\\
&&\hspace{-0.6cm}
\chi^{(3;2B)}_F(t,t_0)=
{1\over2}\int_{t_0}^t\!dt_1H(t_1)^2\int_{t_0}^{t_1}\!dt_2H(t_2)R^{(3;2B)}_F(t,t_1,t_2)
\nonumber\\
&&\hspace{1.6cm}
R^{(3;2B)}_F(t,t_1,t_2)=
\b^3\sum_{k,l,m}F_kL_{km}^{(2)}(t,t_1)L_{ml}^{(1)}(t_1,t_2)p_l(t_2)
\nonumber
\Ee
Finally, the third-order contribution is:
\Be\label{R33.def}
\chi^{(3;3)}_F(t,t_0)
&&\hspace{-0.6cm}=
{1\over6}\int_{t_0}^t\!dt_1H(t_1)^3R^{(3;3)}_F(t,t_1)
\\
&&\hspace{-0.6cm}
R^{(3;3)}_F(t,t_1)=
\b^3\sum_{k,l}F_kL_{kl}^{(3)}(t,t_1)p_l(t_1)
\nonumber
\Ee
These expressions are valid for any Markov process obeying the ME, eq.(\ref{ME.t.abh}) and arbitrary initial conditions (initial populations $p_l(t_0)$).

If the process considered is stationary, meaning that the transition rates are time-independent, 
$W_{kl}(t)=W_{kl}$, the Green's functions depend only on the time-differences, 
$G_{kl}(t_2,t_1)=G_{kl}(t_2-t_1)$.
Furthermore, if the system was in equilibrium initially, $p_l(t_0)=p_l^{\rm eq}$, the expressions simplify considerably.
In this case, the integrals can easily be transformed in order to find expressions for $\chi^{(n)}(t-t_0)$ that are reminiscent of the standard ones used for instance in the field of nonlinear optics\cite{muk95}.
If one is interested in the stationary response, one just starts recording $\chi^{(n)}(t-t_0)$ after times long compared to the initial transients.

Finally, the explicit choice of the field-dependence of the transition rates enters via eq.(\ref{Lkl.def}).
\section*{Appendix B: Nonlinear response functions for the trap model}
\setcounter{equation}{0}
\renewcommand{\theequation}{B.\arabic{equation}}
Using the general expressions given in Appendix A, one can calculate the response for the trap model.
In contrast to other models, the calculation is simplified by the fact that for large $N$, the number of states, one has to consider only terms of order unity and one can neglect all terms of order $1/N$.
In a discrete notation, one has for the trap model $G_{kl}(t)=\d_{kl}e^{-\k_kt}+{\cal O}(1/N)$ with 
$G_{kl}(t)=G(\e_k,t|\e_l)$ and $\k_k=\k(\e_k)$.
Furthermore, the density of state, $\rho_k=\rho(\e_k)$ and the equilibrium populations
$p_k^{\rm eq}=p^{\rm eq}(\e_k)$ scale as $1/N$.
One thus can neglect a number of terms in the calculations.

In the actual calculations, eq.(\ref{Wkl.HX}) is used for the field-dependence of the transition rates.
With this, one has for the relevant part of $L_{kl}^{(\eta)}(t_2,t_1)=L_{kl}^{(\eta)}(t_2-t_1)$ according to 
eq.(\ref{Lkl.def}):
\[
L_{kl}^{(\eta)}(t)=e^{-\k_kt}\k_l\left(\rho_kX_{kl}^\eta-\d_{kl}\overline{X_l^\eta}\right)+{\cal O}(1/N)
\]
where $X_{kl}=\g M_k-\mu M_l$ and $\overline{X_l^\eta}=\sum_k\rho_kX_{kl}^\eta$.
The system is assumed to be in thermal equilibrium in the beginning and the field is assumed to be of the form $H(t)=H_0\cos{(\om t)}$, cf. eq.(\ref{H.cos.om.t}).
The expressions given in Appendix A are used to compute the frequency-dependent response functions according to eq.(\ref{Chi.om.def}).
For the variables $M_k=M(\e_k)$, the choice discussed in the text is used, cf. eqns.(\ref{MkMl.mit}) and
(\ref{Mh4.mit.Gauss}).
Furthermore, one can utilize detailed balance in the form 
\be\label{Det.Bal.Trap}
\rho_k\lg\k\rg=\k_kp_k^{\rm eq}
\quad\mbox{with}\quad
\lg\k\rg=\sum_k\k_kp_k^{\rm eq}
\ee

For the linear response one finds according to eq.(\ref{Chi1})
\be\label{Chi1.Trap}
\chi_1(\om)=(\g+\mu)\b\sum_kp_k^{\rm eq}\lg M_k^2\rg{\k_k\over\k_k-i\om}
\ee
which is just the Fourier transform of the time-derivative of the correlation function $C_M(t)$.
For the third-order response, one finds for $n=1,3$:
\be\label{Chi3.om.Trap}
\chi_3^{(\a)}(\om)={1\over4}\b^3(\g+\mu)
\left\{
\hat\chi_{3;1}^{(\a)}(\om)+\hat\chi_{3;2A}^{(\a)}(\om)+\hat\chi_{3;2B}^{(\a)}(\om)+\hat\chi_{3;3}^{(\a)}(\om)
\right\}
\ee
where the individual terms are given by:
\Be\label{Chi3.1.Trap}
\hat\chi_{3;1}^{(\a)}(\om)=
&&\hspace{-0.6cm}
      3\mu^2\sum_k\rho_k\lg M_k^2\rg^2S_{kkk}^{(\a)}(\om)
      -\mu^2\sum_{k,l}\rho_k\rho_l\lg M_k^2\rg\lg M_l^2\rg S_{kkl}^{(\a)}(\om)
\\
&&\hspace{-0.6cm}
      +\g\mu\sum_{k,l}\rho_k\rho_l\lg M_k^2\rg\lg M_l^2\rg S_{kll}^{(\a)}(\om)
      -\g\mu\sum_{k,l,m}\rho_k\rho_m\rho_l\lg M_k^2\rg\lg M_l^2\rg S_{kml}^{(\a)}(\om)
\nonumber      
\Ee      
with
\Be\label{Sklm.n}
{\rm Re}(S_{klm}^{(1)}(\om))
&&\hspace{-0.6cm}=
\k_m\k_l\lg\k\rg{3\k_k\k_l^2\k_m+\om^2(8\k_k\k_m-2\k_k\k_l-2\k_l\k_m-\k_l^2)
                 \over\k_l(\k_k^2+\om^2)(\k_m^2+\om^2)(\k_l^2+4\om^2)}
\nonumber\\                
{\rm Im}(S_{klm}^{(1)}(\om))
&&\hspace{-0.6cm}=                
\k_m\k_l\lg\k\rg\om{\k_k\k_l^2+2\k_k\k_m\k_l+3\k_l^2\k_m+2\om^2(4\k_m-\k_l)
                    \over\k_l(\k_k^2+\om^2)(\k_m^2+\om^2)(\k_l^2+4\om^2)}
\\                
{\rm Re}(S_{klm}^{(3)}(\om))
&&\hspace{-0.6cm}=
\k_m\k_l\lg\k\rg{\k_k\k_l\k_m-\om^2(2\k_k+3\k_l+6\k_m)
                 \over(\k_m^2+\om^2)(\k_l^2+4\om^2)(\k_k^2+9\om^2)}
\nonumber\\                
{\rm Im}(S_{klm}^{(3)}(\om))
&&\hspace{-0.6cm}=                
\k_m\k_l\lg\k\rg\om{\k_k\k_l+2\k_k\k_m+3\k_l\k_m-6\om^2
                    \over(\k_m^2+\om^2)(\k_l^2+4\om^2)(\k_k^2+9\om^2)}
\nonumber            
\Ee                
This term corresponds to the first line in Fig.\ref{Plot1}.
The second-order terms are:
\Be\label{Chi3.2A.Trap}
\hat\chi_{3;2A}^{(\a)}(\om)=(\g-\mu)
&&\hspace{-0.6cm}
\left\{
       \mu\sum_k\rho_k\lg M_k^2\rg\left(3\lg M_k^2\rg-\overline{\lg M^2\rg}\right) S_{A;kk}^{(\a)}(\om)
\right.      
\\
&&\hspace{-0.3cm}
\left.
      +\g\sum_{k,l}\rho_k\rho_l\lg M_k^2\rg\left(\lg M_k^2\rg-\overline{\lg M^2\rg}\right) S_{A;kl}^{(\a)}(\om)
\right\}      
\nonumber
\Ee      
where the corresponding spectral functions are:
\Be\label{SAkl.n}
{\rm Re}(S_{A;kl}^{(1)}(\om))
&&\hspace{-0.6cm}=
\lg\k\rg{1\over2}{3\k_k\k_l^2+2\om^2(4\k_k-\k_l)\over(\k_k^2+\om^2)(\k_l^2+4\om^2)}
\nonumber\\                
{\rm Im}(S_{A;kl}^{(1)}(\om))
&&\hspace{-0.6cm}=                
\lg\k\rg{1\over2}\om{3\k_l^2+2\k_k\k_l+8\om^2\over(\k_k^2+\om^2)(\k_l^2+4\om^2)}
\\                
{\rm Re}(S_{A;kl}^{(3)}(\om))
&&\hspace{-0.6cm}=
\k_l\lg\k\rg{1\over2}{\k_k\k_l-6\om^2\over(\k_l^2+4\om^2)(\k_k^2+9\om^2)}
\nonumber\\                
{\rm Im}(S_{A;kl}^{(3)}(\om))
&&\hspace{-0.6cm}=                
\k_l\lg\k\rg{1\over2}\om{2\k_k+3\k_l\over(\k_l^2+4\om^2)(\k_k^2+9\om^2)}
\nonumber            
\Ee                
Additionally, I defined 
\be\label{Mh2.mit.Trap}
\overline{\lg M^2\rg}=\sum_k\rho_k\lg M_k^2\rg
\ee
The other second-order term is:
\Be\label{Chi3.2B.Trap}
\hat\chi_{3;2B}^{(\a)}(\om)=
&&\hspace{-0.6cm}
-3\mu^2\sum_k\rho_k\lg M_k^2\rg^2S_{B;kk}^{(\a)}(\om)
-\g^2\overline{\lg M^2\rg}\sum_k\rho_k\lg M_k^2\rg S_{B;kk}^{(\a)}(\om)
\\
&&\hspace{-0.6cm}
-2\g\mu\sum_{k,l}\rho_k\rho_l\lg M_k^2\rg\lg M_l^2\rg S_{B;kl}^{(\a)}(\om)
\nonumber
\Ee      
with
\Be\label{SBkl.n}
{\rm Re}(S_{B;kl}^{(1)}(\om))
&&\hspace{-0.6cm}=
\k_l\lg\k\rg{1\over2}{3\k_k\k_l-\om^2\over(\k_k^2+\om^2)(\k_l^2+\om^2)}
\nonumber\\                
{\rm Im}(S_{B;kl}^{(1)}(\om))
&&\hspace{-0.6cm}=                
\k_l\lg\k\rg{1\over2}\om{\k_k+3\k_l\over(\k_k^2+\om^2)(\k_l^2+\om^2)}
\\                
{\rm Re}(S_{B;kl}^{(3)}(\om))
&&\hspace{-0.6cm}=
\k_l\lg\k\rg{1\over2}{\k_k\k_l-3\om^2\over(\k_l^2+\om^2)(\k_k^2+9\om^2)}
\nonumber\\                
{\rm Im}(S_{B;kl}^{(3)}(\om))
&&\hspace{-0.6cm}=                
\k_l\lg\k\rg{1\over2}\om{\k_k+3\k_l\over(\k_l^2+\om^2)(\k_k^2+9\om^2)}
\nonumber            
\Ee                
Finally, the term corresponding to the third-order perturbation, i. e. the last line in Fig.\ref{Plot1}, is given by:
\be\label{Chi3.3.Trap}
\hat\chi_{3;3}^{(\a)}(\om)=
3(\mu^2-\g\mu+\g^2)\sum_k\rho_k\lg M_k^2\rg^2S_k^{(\a)}(\om)
+3\g\mu\overline{\lg M^2\rg}\sum_k\rho_k\lg M_k^2\rg S_k^{(\a)}(\om)
\ee  
where
\Be\label{Sk.n}
{\rm Re}(S_k^{(1)}(\om))
&&\hspace{-0.6cm}=
\lg\k\rg{1\over2}{\k_k\over(\k_k^2+\om^2)}
\quad;\quad
{\rm Im}(S_k^{(1)}(\om))
=\lg\k\rg{1\over2}{\om\over(\k_k^2+\om^2)}
\\                
{\rm Re}(S_k^{(3)}(\om))
&&\hspace{-0.6cm}=
\lg\k\rg{1\over6}{\k_k\over(\k_k^2+9\om^2)}
\quad;\quad
{\rm Im}(S_k^{(3)}(\om))
=\lg\k\rg{1\over2}{\om\over(\k_k^2+9\om^2)}
\nonumber            
\Ee                
Here, all expressions are given in a discrete notation.
If one changes to a continuous description, one has to replace all sums by the appropriate integrals over the trap energies $\e_k$.
\end{appendix}
\newpage

%
\end{document}